\documentclass[10pt]{iopart}
\begin{document}

\title{Next-to-next-to-leading order fits to  CCFR'97 $xF_3$ data 
and infrared renormalons}  
 

\author{A.L. Kataev\dag , G. Parente \ddag and A.V. Sidorov$^{*}$ 
}

\address{\dag\ Institute for Nuclear Research of the Academy 
of Sciences of Russia, 117312, Moscow, Russia} 

\address{\ddag\ Department of Particle Physics, University of Santiago
de Compostela, 15706 Santiago de Compostela, Spain}

\address{$^{*}$ Bogoliubov Laboratory of Theoretical Physics, 
Joint Institute for Nuclear Research, 141980, Russia}

\ead{kataev@ms2.inr.ac.ru, gonzalo@fpaxp1.usc.es, sidorov@thsun1.jinr.ru}

\begin{abstract}
We briefly summarize the outcomes 
of our recent  improved fits to the  experimental data 
of CCFR collaboration for  $xF_3$ structure function of  
$\nu N$ deep-inelastic scattering at the next-to-next-to-leading order.
Special attention is paid to the extraction of $\alpha_s(M_Z)$ and 
the parameter of the infrared  renormalon model for $1/Q^2$-correction at 
different orders of perturbation theory. The results can be of interest for 
planning similar studies using possible future data 
of Neutrino Factories.   
\end{abstract}





The most precise up to date data for  $\nu N$
deep-inelastic scattering cross-sections  were  provided in 1997 by 
CCFR collaboration \cite{Seligman:mc}. In view of the necessity of 
understanding of the place of different theoretical effects in the 
fits to $xF_3$ non-singlet structure function data, interesting from the 
point of view of possible future theoretical program of Neutrino Factories 
(for a review see Ref.\cite{Mangano:2001mj}), it is worth to summarize 
some basic results of the most recent higher-order perturbative QCD 
analysis \cite{Kataev:2001kk}, which updates the results of the 
previous similar studies of Refs.\cite{Kataev:1997nc},\cite{Kataev:1999bp}.
The analysis is based on the Jacobi polynomial technique of reconstruction 
of the structure function from its Mellin moments
\begin{equation}
xF_3(x,Q^2)=w(\alpha,\beta)\sum_{n=0}^{N_{max}}
\Theta_n^{\alpha,\beta}(x)\sum_{j=0}^{n}c_j^{(n)}(\alpha,\beta)M_{j+2}^{TMC}
(Q^2)+\frac{h(x)}{Q^2}
\end{equation}
where $w(\alpha,\beta)=x^{\alpha}(1-x)^{\beta}$, $\Theta_n^{\alpha,\beta}$
are the Jacobi polynomials, $c_j^{(n)}(\alpha,\beta)$ contain 
$\alpha-$ and $\beta-$ Euler $\Gamma$-functions with the Jacobi polynomial 
parameters $\alpha=0.7$ and $\beta=3$ fixed following the detailed studies 
of Ref.\cite{Kataev:1999bp}.

The moments 
\begin{equation}
M_n^{TMC}(Q^2)=M_n(Q^2)+\frac{n(n+1)}{(n+2)}
\frac{M_{nucl}^2}{Q^2}M_{n+2}(Q^2)
\end{equation}
contain the leading order target mass 
correction, while the contribution of higher  terms of order 
$1/Q^4$ do not affect the results of the fits \cite{Kataev:1999bp}.
Here we will fix the dynamical  non-perturbative corrections 
using infrared renormalon (IRR) model as 
\begin{equation}
\frac{h(x)}{Q^2}=x^{\alpha}(1-x)^{\beta}\sum_{n=0}^{N_{max}}
\Theta_n^{\alpha,\beta}\sum_{j=0}^{n}c_j^{(n)}(\alpha,\beta)M_{j+2}^{IRR}(Q^2)
\end{equation}
where $M_n^{IRR}(Q^2)=\tilde{C}(n)M_n(Q^2)A_2^{'}/Q^2$, 
$\tilde{C}(n)=-n+4+2/(n+1)+4(n+2)+4S_1(n)$ was calculated from the 
single chain of one-fermion loop insertions into the one- gluon 
contribution to the Born diagram  in Ref.\cite{Dasgupta:1996hh} 
and $A_2^{'}$ is the arbitrary 
fitted parameter.

The Mellin moments obey the following renormalization-group 
equation 
\begin{equation}
\frac{M_n(Q^2)}{M_n(Q_0^2)}=\rm exp\bigg[-\int_{A_s(Q_0^2)}^{A_s(Q^2)}
\frac{\gamma_{F_3}^{(n)}(\eta)}{\beta(\eta)}d\eta\bigg]\frac{C_{F_3}^{(n)}(A_s(Q^2))}
{C_{F_3}^{(n)}(A_s(Q_0^2))}
\end{equation} 
where $A_s=\alpha_s/(4\pi)$, while $M_n(Q_0^2)$ is the phenomenological 
quantity, defined at the initial scale $Q_0^2$ as $M_n(Q_0^2)=\int_0^1
x^{n-2}A(Q_0^2)x^b(Q_0^2)(1+\gamma(Q_0^2))dx$.

The main new ingredient of the analysis of Ref.\cite{Kataev:2001kk}
comes from the recent results of analytical calculations of 
next-to-next-to-leading order (NNLO) 
correction to $\gamma_{F_3}^{(n)}(A_s)$ at odd $n=3,5,7,...,13$ 
\cite{Retey:2000nq} (the NNLO corrections to $\beta(x)$ \cite{Tarasov:au}
and $C_{F_3}^{(n)}(A_s)$ \cite{Zijlstra:1992kj} were already known 
for quite a long time). The switch from the NNLO non-singlet 
contributions 
to  $\gamma_{F_2}^{(n)}(A_s)$ at n=2,4,..,10 calculated  
in  Ref.\cite{Larin:1993vu}, which numerically are closely related 
to the identical contributions to $\gamma_{F_3}^{(n)}(A_s)$,  and  
were used in the previous NNLO fits
of CCFR'97 data of Refs.\cite{Kataev:1997nc},\cite{Kataev:1999bp}, is making 
our new analysis more self-consistent. Note, however, that in spite  
of fixation of the number of previously existing theoretical 
ambiguities, our new fits  are   confirming  
the main results of Ref.\cite{Kataev:1999bp}, putting them on more 
solid ground.  

In order to save space, we will avoid discussions of the number of 
technical steps of the analysis of Ref.\cite{Kataev:2001kk} and 
will concentrate ourselves on the presentation of the main physical 
results. 

\begin{center}
\begin{tabular}{||c|c|c|c|c||}
\hline
order/$N_{max}$ &  $Q_0^2=$&
5~Gev$^2$& 20~GeV$^2$  & 100~GeV$^2$ \\ \hline   

LO/9 & $\Lambda_{\overline{MS}}^{(4)}$ & 447$\pm$54 & 443$\pm$54 &439$\pm$56\\
       &$\chi^2$/nep & 79.8/86 & 80.1/86 & 79.6/86 \\ 
       &$A_2^{'}$ & $-$0.340$\pm$0.059 & $-$0.337$\pm$0.059 & $-$0.335$\pm$0.059 \\
\hline

NLO/9 & $\Lambda_{\overline{MS}}^{(4)}$ & 379$\pm$41 & 376$\pm$39 &374$\pm$42\\
       &$\chi^2$/nep & 78.6/86 & 79.5/86 & 79.0 \\ 
       &$A_2^{'}$ & $-$0.125$\pm$0.053 & $-$0.125$\pm$0.053 & $-$0.124$\pm$0.053 \\

\hline

NNLO/9 & $\Lambda_{\overline{MS}}^{(4)}$ & 331$\pm$33 & 332$\pm$35 &331$\pm$35
\\
       &$\chi^2$/nep & 73.1/86 & 75.7/86 & 76.9/86 \\ 
       &$A_2^{'}$ & $-$0.013$\pm$0.051 & $-$0.015$\pm$0.051 & $-$0.016$\pm$0.051 \\
\hline
\end{tabular}
\end{center}
{{\bf Table 1.} The results of the fits to  the CCFR'97   $xF_3$ data 
with $Q^2>5~GeV^2$ with related statistical uncertainties.
The cases of  different $Q_0^2$  are considered. The values of 
$\Lambda_{\overline{MS}}^{(4)}$ are given in MeV, while $A_2^{'}$ 
is ``measured'' in GeV$^2$.} 

In Table 1 we present the results of the fits taking into 
account leading order (LO), next-to-leading order (NLO) 
and NNLO perturbative QCD corrections. 
Due to the incorporation 
of the recently calculated NNLO corrections to $\gamma_{F_3}^{(n)}(\alpha_s)$
\cite{Retey:2000nq}  
into our new NNLO  fits of Ref.\cite{Kataev:2001kk}
the NNLO fitted parameters turn out to be rather stable 
to the variation
 of $Q_0^2$-scale from 20 GeV$^2$, 
used in Ref.\cite{Kataev:1999bp}, to the lower initial scale $Q_0^2=5$ GeV$^2$.
It should be stressed, that 
the tendency $|A_2^{'} |_{LO}>|A_2^{'} |_{NLO}>|A_2^{'} |_{NNLO}$  
demonstrates the minimization of the role  of the  power corrections
for the description of $xF_3$ data.
Moreover, at the NNLO the value of the IRR model  parameter 
$A_2^{'}$  is effectively minimized and within statistical 
uncertainties is comparable with zero.
This property was already observed in the process of our previous 
NNLO analysis of CCFR'97 data  \cite{Kataev:1997nc,Kataev:1999bp} and 
in the NNLO fits to the combined data of SLAC,NMC and BCDMS collaboration for 
$F_2$ structure function of deep-inelastic scattering of charged 
leptons on nucleons \cite{Yang:1999xg}.In all these 
fits the QCD coupling 
constant was closed to its world average value $\alpha_s(M_Z)\approx 0.118$.
For example, in our analysis of Ref.\cite{Kataev:2001kk} we got 
\begin{eqnarray}
NLO~\alpha_s(M_Z)&=&0.120 \pm 0.002(stat)
\pm 0.005(syst) \\ \nonumber
&&\pm 0.002(thresh.)^{+0.010}_{-0.006}(scale)
\end{eqnarray}
\begin{eqnarray}
NNLO~\alpha_s(M_Z)&=&0.119\pm 0.002(stat)
\pm 0.005(syst) \\ \nonumber
&&\pm 0.002(thresh.)^{+0.004}_{-0.002}(scale)
\end{eqnarray}
The latter result agree, within errors,  with the results of the NNLO 
fits to CCFR'97 data, performed with the help of 
 Bernstein polynomial technique using different theoretical 
representations (compare the original publication of 
Ref.\cite{Santiago:2001mh} with the recent work of Ref.\cite{Maxwell:2002mt}). 
Note, however, that if one gets the ``small''  NNLO value of $\alpha_s(M_Z)$,
namely  $\alpha_s(M_Z)\approx 0.114$, like in the work of 
Ref.\cite{Alekhin:2002wp}, the $1/Q^2$  contributions
to $F_2$ remain visible even at the NNLO \cite{Alekhin:2002wp}.
Thus, we can conclude, that there is some correlation between twist-4 
terms and NNLO perturbative QCD corrections and NNLO value of  $\alpha_s(M_Z)$.
One can hope, that the 
possibility to detect higher-twist terms at the NNLO 
will be clarified in future 
using drastically more precise data for $F_2$ and $xF_3$ 
structure functions, 
which can be obtained at Neutrino Factories in the lower $Q^2$-region.

{\bf Acknowledgments} 

The work of two of us (ALK and AVS) was supported by RFBR Grants N 00-02-17432
and N 02-01-00601. 
GP acknowledges the support of Xunta de Calicia ( PGIDT00PX20615PR) 
and CICYT (AEN99-0589-C02-02).
AVS is grateful to INTAS (INTAS Call 2000-project N 587)
for the financial support. 
One of us (ALK) wishes to thank the OC of NuFact'02 Workshop for 
hospitality in London.

\end{document}